\documentclass[12pt]{article}
\usepackage[totalheight = 23cm, totalwidth = 17cm]{geometry}
\newcommand{\nin}{\noindent}
\newcommand{\be}{\begin{equation}}
\newcommand{\ee}{\end{equation}}
\newcommand{\bea}{\begin{eqnarray}}
\newcommand{\eea}{\end{eqnarray}}
\newcommand{\br}{\hskip .25cm/\hskip -.25cm}

\newcommand{\nn}{\nonumber\\}

\newcommand{\ol}{\overline}
\usepackage{graphicx}

\begin{document}

\begin{center}

{\bf{\Large Branon stabilization from fermion-induced radiative corrections}}

\vspace{1cm}

J. Alexandre\footnote{jean.alexandre@kcl.ac.uk} and D. Yawitch\footnote{darrel.yawitch@kcl.ac.uk}\\
Department of Physics, King's College London, WC2R 2LS, UK

\vspace{1cm}

{\bf Abstract}

\end{center}

\nin 
We consider a 3-brane embedded in 5-dimensional space time characterized by a Gaussian warp factor, 
for which the four-dimensional effective theory for brane fluctuations (branons) is unstable. 
We show that radiative corrections arising from fermions living 
on the brane, and therefore coupled to branons, stabilize the system by generating dynamically a spontaneous 
symmetry breaking for the 
branon field. The price to pay, for the corresponding mechanism to be consistent, is to have a 
large number of fermion flavours,
and we discuss the fat brane scenario as an interpretation for the dressed branon theory,
taking into account the Maxwell construction, which avoids the spinodal instability present in the perturbative effective
potential.

\vspace{1cm}

{\it Introduction.}
In the context of higher dimensional models, Brane Cosmology provides relations between bulk quantities and observables
on the brane, which leads to possible explanations of mass hierarchy (reviews 
on Brane Cosmology can be found for example in \cite{review}).
In this framework, fluctuations of the position of a four-dimensional brane in a higher-dimensional bulk can lead to an 
effective four-dimensional theory, where branons (particles associated to these fluctuations) are described by a 
scalar field living on a flat brane, and correspond to Goldstone bosons associated to the
spontaneous symmetry breaking of translations in the transverse direction, 
as a consequence of the presence of the brane \cite{dvali}.
This idea has been developed in \cite{yochioka}, where the authors show how 
these brane transverse degrees of freedom suppress the coupling of Kaluza-Klein modes to matter on the brane. 
The relevance of branons to dark matter, 
and the study of their coupling to particles of the Standard Model is studied in \cite{dobado}. \\
In the present work, we study  the effects of a specific warp factor on branon fluctuations.
The warp factor, which is even in the extra coordinate $y$, is assumed to be differentiable,  
in order to define a quantum theory for fluctuations, using the usual path integral quantization.
We consider therefore a Gaussian warp factor, for which the branon system on its own is not stable.
Nevertheless, we show that stability of branons can be restored, by radiative corrections 
arising from fermions living on the brane. The corresponding process is similar to the 
Coleman-Weinberg mechanism \cite{coleman}, and we find that, for
the dynamical generation of spontaneous symmetry breaking to be consistent in this scenario,
few hundreds of fermion flavours are necessary. 
This situation is not in ruled out, as there is no experimental constraint on the number of right-handed neutrinos,
and the influence of a large number of flavours on the seesaw mechanism is discussed in \cite{ellis}. 
We discuss then the spinodal instability induced by the non-convex one-loop effective potential for branons, 
and explain how the Maxwell construction saves us from this perturbative instability. 
The resulting vacuum for branons is made of condensates for the branon field, 
and corresponds to a shifted coordinate for the
ground state of the brane, which can seem inconsistent if the warp factor is expected to be centered on the brane. 
We note here that this is not necessarily the case, as was shown in \cite{probe}, where a probe brane (with 
a tension and a charge both going to zero, but keeping a fixed ratio) moves away from the maximum of the 
warp factor, with a typical time scale of the order of the age of the Universe, such that the 
corresponding instability is acceptable.
This scenario, though, involves two other branes, one of which is on the maximum of the warp factor, and
we will assume here one brane only in the vicinity of the maximum of the warp factor, and with a finite tension.
As an interpretation of our results, a consistent picture is provided by a fat brane interpolating
the two non-trivial minima of the one-loop effective potential for branons. 
A fat brane is expected with the initial choice of a 
Gaussian warp factor, and we show that the corresponding brane thickness is given by $2/f$ at most, where $f$ is the cut
off provided by the brane tension $f^4$.

{\it Classical effective action for the branon field.}
We consider a 5-dimensional Universe with generic coordinates $X^M=(x^\mu,y)$, where $x$ are the coordinates on the brane,
which is defined by the equation $y=Y(x)$. In what follows, the brane coordinates are denoted with the indices
$\mu,\nu$ and the bulk coordinates with $M,N$. 
Motivated by scenarios involving confinement on the brane, we consider the following block-diagonal bulk metric 
\be
g_{MN}=\left(\begin{array}{cc}e^{2\sigma(y)}\eta_{\mu\nu}&0\\0&-1\end{array}\right),
\ee
where the warp factor we consider is assumed to be even in $y$, and differentiable. We will take here
\be\label{sigma}
\sigma(y)=-\frac{m^2}{2}y^2,
\ee
where $m$ is a mass scale of the theory, as was done in \cite{gogberashvili}. The induced metric $h_{\mu\nu}$ on the brane is 
\be\label{induced}
h_{\mu\nu}(x)=\partial_\mu X^M\partial_\nu X^N g_{MN}(X)=e^{2\sigma(Y)}\eta_{\mu\nu}-\partial_\mu Y\partial_\nu Y,
\ee
and, if $f^4$ is the brane tension, the brane action is then 
\bea
S_{brane}&=&-f^4\int d^4x\sqrt{-h}\\
&=&-f^4\int d^4x~e^{4\sigma(Y)}\left( 1-\frac{1}{2}e^{-2\sigma(Y)}\eta^{\mu\nu}\partial_\mu Y\partial_\nu Y
+\cdots\right), \nonumber
\eea
where dots represent higher orders in derivatives of $Y$, which will be disregarded in the framework 
of the gradient expansion.\\
Our dynamical variable is the canonically normalized branon field $\phi=f^2Y$, with mass dimension 1, and
the classical brane ground state is $Y=0$.
The resulting effective action for branons is then
\be\label{Seff}
S_{eff}=\int d^4x\left( \frac{e^{2\sigma(\phi)}}{2}\partial^\mu\phi\partial_\mu\phi-f^4e^{4\sigma(\phi)}\right),
\ee
and contains derivative and polynomial interactions. In \cite{dobado}, the authors consider a general metric
$\tilde g_{\mu\nu}$ for the restriction of the bulk metric on the brane, 
and expand the former in powers of $y$, which leads to a 
mass term for the branon field. In our case, though, the Gaussian warp factor does not allow any restoration force 
arising from the potential, and induces an instability. Indeed, the branon classical mass squared $M_B^2$ is defined 
as the pole of the propagator of fluctuations around the 
minimum of the potential, which, with the warp factor (\ref{sigma}), arises at $\phi_{min}=\infty$ and gives
\be 
M_B^2=e^{-2\sigma(\phi_{min})}\times 4f^4\Big(\sigma''(\phi_{min})+4[\sigma'(\phi_{min})]^2\Big)e^{4\sigma(\phi_{min})}=0.
\ee

{\it One-loop effective potential.}
We now consider the one-loop theory arising from the classical action (\ref{Seff}).
As expected from radiative corrections arising from self-interacting bosons, quantum effects generated 
by the theory (\ref{Seff}) will not cure the stability problem, as we show now.
Quantum effects generate, among others, purely polynomial interactions, independent of derivative of the branon field. 
Indeed, the one-loop effective potential obtained from the classical theory (\ref{Seff}) is given by
\be
U^{(1)}(\phi)=f^4e^{4\sigma(\phi)}+
\frac{1}{2V}\mbox{Tr}\left\lbrace\ln\left(\frac{\delta^2S_{eff}}{\delta\phi\delta\phi}\right)\right\rbrace, 
\ee
where $V$ is the brane volume and the second derivative of the action (\ref{Seff}) 
is taken for the constant configuration (in Euclidean space):
\be
\left.\frac{\delta^2S_{eff}}{\delta\phi(x)\delta\phi(y)}\right|_{const.}
=e^{2\sigma}\left(p^2+4f^4e^{2\sigma}[\sigma''+4(\sigma')^2]\right)\delta^{(4)}(x-y).
\ee 
If we ignore field-independent terms and 
take the momentum cut off $f$, the one-loop effective potential is then given by
\be
f^{-4}U^{(1)}(\phi)=e^{4\sigma}+\frac{\sigma}{32\pi^2}+\frac{1}{64\pi^2}
\left[ \Delta+\ln(1+\Delta)-\Delta^2\ln(1+\Delta^{-1})\right] ,
\ee
where $\Delta=4f^2e^{2\sigma}[\sigma''+4(\sigma')^2]$.
In the case where $\sigma$ is given by eq.(\ref{sigma}), 
we obtain, in terms of the dimensionless parameter $a=m^2/f^2$ and dimensionless field $\xi=\phi/f$,
\bea\label{U1}
f^{-4}U^{(1)}(\xi)&&=e^{-2a\xi^2}-\frac{a\xi^2}{64\pi^2}\nn
&&+\frac{1}{64\pi^2}\Bigg[4ae^{-a\xi^2}(4a\xi^2-1)+\ln\left(1+4ae^{-a\xi^2}(4a\xi^2-1)\right)\nn
&&~~~~~~~~-16a^2e^{-2a\xi^2}(4a\xi^2-1)^2\ln\left(1+e^{a\xi^2}\big(16a^2\xi^2-4a\big)^{-1}\right)\Bigg].
\eea
This one-loop potential is, in terms of stability, worse than the classical one, and 
we shall add fermions on the brane, in order to stabilize branons, since the one-loop fermionic
contribution comes with an opposite sign.

{\it Coupling to fermions.}
In what follows, Latin indices refer to the local inertial frame and are contracted with $\eta_{ab}$,
whereas Greek indices are contracted with the induced metric $h_{\mu\nu}$.\\
Neglecting higher orders in the branon derivatives $\partial Y$, we find the following approximate vierbeins on the brane
\be
e_\mu^a\simeq\delta_\mu^a~ e^{\sigma(Y)}-\frac{e^{-\sigma(Y)}}{2}\partial_\mu Y\partial^a Y,
\ee
which lead to the expected definition $e_\mu^a e_\nu^b\eta_{ab}=h_{\mu\nu}$, up to
higher orders in $\partial Y$. We have then
\be
e_a^\mu\simeq\delta_a^\mu~ e^{-\sigma(Y)}+\frac{e^{-3\sigma(Y)}}{2}\partial_a Y\partial^\mu Y
\ee
and the total action for branons and $N$ fermions with masses $\tilde m_i$ is \cite{birrell}
\be\label{Stotal}
S_{total}=-f^4\int d^4x\sqrt{-h}+\sum_{i=1}^N\int d^4x
\sqrt{-h}\left\lbrace \frac{i}{2}\ol\psi_i\gamma^ae_a^\mu{\cal D}_\mu\psi_i
-\frac{i}{2}e_a^\mu\left({\cal D}_\mu\ol\psi_i\right)\gamma^a\psi_i -\tilde m_i\ol\psi_i\psi_i\right\rbrace ,
\ee
where ${\cal D}_\mu=\partial_\mu+\Gamma_\mu$ and the spin connection is 
\be
\Gamma_\mu=\frac{1}{8}[\gamma^a,\gamma^b]e_a^\nu\nabla_\mu e_{b\nu}.
\ee
The gamma matrices $\gamma^a$ are defined in the local inertial frame, and therefore satisfy
\be
\{\gamma^a,\gamma^b\}=2\eta^{ab}.
\ee
The fermion sector of the action (\ref{Stotal}) is then, for a constant branon configuration, 
\be\label{Sefftilde}
\tilde S_{eff}=\sum_{i=1}^N\int d^4x~e^{3\sigma}~\ol\psi_i\left( i\br\partial-\tilde m_ie^{\sigma}\right)\psi_i ,
\ee
and the corresponding one-loop effective potential for branons is then 
\bea
\tilde U^{(1)}(\phi)&=&
-\frac{1}{V}\sum_{i=1}^N
\mbox{Tr}\left\lbrace\ln\left(\frac{\delta^2\tilde S_{eff}}{\delta\psi_i\delta\ol\psi_i}\right) \right\rbrace\nn
&=&-\frac{1}{2V}\sum_{i=1}^N
\mbox{Tr}\left\lbrace\ln\left[\left(\frac{\delta^2\tilde S_{eff}}{\delta\psi_i\delta\ol\psi_i}\right)
\left(\frac{\delta^2\tilde S_{eff}}{\delta\psi_i\delta\ol\psi_i}\right)^\dagger\right]\right\rbrace\nn
&=&-\frac{1}{2V}\sum_{i=1}^N
\mbox{Tr}\left\lbrace\ln\left[e^{6\sigma}(p^2+e^{2\sigma}\tilde m_i^2)\delta^{(4)}(x-y)\right]\right\rbrace,
\eea
where a Euclidean metric is used. We obtain then, in terms of the dimensionless parameters $b_i=\tilde m_i^2/f^2$,
\be\label{U1tilde}
f^{-4}\tilde U^{(1)}(\phi)=-\frac{3N\sigma}{32\pi^2}+\frac{1}{64\pi^2}\sum_{i=1}^N\left[-b_ie^{2\sigma}
-\ln(1+b_ie^{2\sigma})+b_i^2e^{4\sigma}\ln(1+b_i^{-1}e^{-2\sigma})\right] 
\ee
where constant terms are discarded.\\
Taking into account the warp factor (\ref{sigma}), the total one-loop effective potential 
$V=U^{(1)}+\tilde U^{(1)}$ is given by
\bea\label{V}
f^{-4}V(\xi)&=&e^{-2a\xi^2}+(3N-1)\frac{a\xi^2}{64\pi^2}\\
&&+\frac{1}{64\pi^2}\Bigg[4ae^{-a\xi^2}(4a\xi^2-1)+\ln\left(1+4ae^{-a\xi^2}(4a\xi^2-1)\right)\nn
&&~~~~~~~~-16a^2e^{-2a\xi^2}(4a\xi^2-1)^2\ln\left(1+[4ae^{-a\xi^2}(4a\xi^2-1)]^{-1}\right)\Bigg]\nn
&&+\frac{1}{64\pi^2}\sum_{i=1}^N\left[
-b_ie^{-a\xi^2}-\ln(1+b_ie^{-a\xi^2})+b_i^2e^{-2a\xi^2}\ln(1+b_i^{-1}e^{a\xi^2})\right].\nonumber
\eea

{\it Spinodal instability.} The result (\ref{V}) shows an instability of the system for certain values of the field and 
the parameters $N,a$, 
as the logarithms can develop imaginary parts when their arguments are negative. This is an artifact, though \cite{nonconvex},
as the effective potential of a theory must be convex \cite{convex}. What is actually happening in the spinodal region 
(the would-be instable region), is the famous Maxwell construction: the full effective potential interpolates the
minima of the potential with a straight line. This mechanism is highly non-perturbative and corresponds to a tree-level
renormalization \cite{tree}. It has been taken into account using Wilsonian renormalization group equations \cite{Maxwell}, 
and the evolution of the coarse grained potential, from the spontaneous symmetry breaking bare potential to the IR convex potential,
was explicitly shown in \cite{induced}. As a consequence, the correct effective potential for branons is flat between the 
minima $\pm\xi_{min}$ of the potential (\ref{V}), and no ill-defined logarithm occurs. 
To find $\xi_{min}$, it is enough to consider the first line of eq.(\ref{V}) only: {\it (i)}
the decreasing exponentials in the other two lines quickly lead the different terms to vanish when the field increases;
{\it (ii)} the factor $3N-1$, as we will see, is of the order $10^3$, dominating over the other terms.
We obtain then the following constraints
\begin{itemize}

\item  We have to check that, outside the spinodal region, the logarithms in eq.(\ref{V}) do not develop 
imaginary parts\footnote{ we thank the referee for pointing this out}, for any value of $\xi$ such that $\xi^2\ge\xi_{min}^2$:
for the logarithms on lines 2 and 3 to be simultaneously real, it is necessary that $4a\xi^2>1$. 
Together with $\xi^2\ge\xi_{min}^2$, this is satisfied if
\be\label{allowed}
N\le \frac{1}{3}+\frac{128\pi^2}{3\sqrt{e}}\simeq 255.
\ee

\item if $N<(128\pi^2+1)/3\simeq 421$, and therefore for the allowed values (\ref{allowed}) of $N$,
the potential(\ref{V}) has two non-vanishing minima defined by
\be\label{ximin}
\xi_{min}^2\simeq\frac{1}{a}\ln\left(\frac{8\pi\sqrt 2}{\sqrt{3N-1}}\right),
\ee
and, although the branon doesn't feel any restoration force between the minima $\pm\xi_{min}$ because the 
potential is flat,
one can still define its mass at the border of the spinodal region, where the potential (\ref{V}) is stable and provides 
a restoration force.
For the sake of clarity, we  again take into account the first line of eq.(\ref{V}) only, and 
the dimensionless branon mass $\mu=M/f$ is then given by
\be
\mu^2=e^{a\xi_{min}^2}f^{-4}\left.\frac{\partial^2 V}{\partial\xi^2}\right|_{\xi_{min}}
\simeq\frac{a}{\pi}\sqrt{2(3N-1)}\ln\left(\frac{8\pi\sqrt 2}{\sqrt{3N-1}}\right).
\ee
In this last expression, $e^{a\xi^2_{min}}$ is the inverse of the wave function normalization, appearing in the branon kinetic term.
A consistent solution for branons stabilization should satisfy
$\xi_{min}^2<1$ and $\mu^2<1$, which corresponds to having the mass and amplitude of the branon field
smaller than the cut off $f$ provided by the brane tension. This imposes the following bounds for $a$ 
\be\label{bounds}
\ln\left(\frac{8\pi\sqrt 2}{\sqrt{3N-1}}\right) < a < \frac{\pi}{\sqrt{2(3N-1)}}
\left[ \ln\left(\frac{8\pi\sqrt 2}{\sqrt{3N-1}}\right)\right]^{-1},
\ee
which can be satisfied for $N\ge 237$ only, for the inequalities (\ref{bounds}) to make sense.
\end{itemize}
Finally, our results are consistent with any number of flavours satisfying
\be\label{range}
237\le N\le 255,
\ee 
and a ratio $a=m^2/f^2$ smaller than one, since for the range (\ref{range}), the bounds (\ref{bounds}) lead to
\be
0.25<a<0.32.
\ee 
As a consequence, provided we have an appropriate number of fermion flavours, 
we can define a consistent field theory for branons. We still have to interpret our result, though, 
in terms of the position of the brane, which is expected to be centered on the maximum of the warp factor.

{\it Fat brane.} 
The initial warp factor (\ref{sigma}) we consider, introduced in the context of fat brane, is consistent with our picture 
for the following reason.
As far as the vacuum is concerned, because of the flat potential and the absence of restoration force, 
condensates with any amplitude between $\pm\xi_{min}$ are allowed, which 
naturally leads to the concept of fat brane since the position of the brane is not defined precisely 
within the interval $[-\xi_{min};\xi_{min}]$.

We note that another scenario involving the Coleman-Weinberg mechanism
was studied in \cite{zarembo}, where the author shows that  
the integration of gauge fields on  parallel D-branes generates dynamically 
a non-trivial vacuum for scalar fields on the branes. 
This leads to the conclusion that a stack of D-branes put on top of each other is unstable, 
but becomes stable when the branes are separated and 'fall' into the minima of the effective potential. 
Such a scenario leads to the concept of fat branes, which can be obtained from a collection of parallel D-branes, 
after integrating out the string interactions among them, as was discussed in \cite{LMZ}. 
Taking into account \cite{zarembo}, even with several branes, our effective potential for branons is then expected to 
keep its symmetry breaking feature at the tree level, thus leading to the Maxwell construction for the non-perturbative
effective potential.
Also, the separation $d=2\xi_{min}/f$ between the two minima of the effective potential (\ref{V}) 
satisfies $d<2/f$, which is consistent with the expected brane thickness, if we assume $f$ to be of the order of the TeV.

{\it Conclusion.}
We showed that the branon model arising from a brane sitting on a Gaussian warp factor 
can be stabilized by a large number of fermion flavours, and is consistent with a fat brane picture.
In this context, dark matter could correspond to quantum fluctuations of the branon field. 
A natural continuation of this work deals with the possible backreaction of the fat brane on the warp factor 
outside the spinodal region, and
we expect that this would lead to mild corrections, since we found that the 
thickness of the brane $\sim 1/f$ is smaller than the thickness of the warp factor $\sim 1/m$. Also,
a similar study should be done for a non-differentiable warp factor, as in the Randall-Sundrum case. This study would
need an extension of the usual path integral quantization approach, though, which would take into account the non-differentiable interaction, and this is left for a future work.

\vspace{0.5cm}

\nin {\bf Acknowledgments}
J.A. would like to thank Janos Polonyi for useful comments. 
The work of D.Y. is supported by the STFC, UK.

\end{document}